\documentclass[12pt]{article}
\usepackage{amsmath}
\usepackage{amssymb}
\usepackage{amsfonts}
\usepackage{amscd}
\usepackage{delarray}
\usepackage{graphicx}

\usepackage{amsmath,amsmath,amsfonts,amssymb,color,bbm}

\usepackage{graphicx}

\def\Symp#1,#2,#3,#4.{\left[\left(\begin{array}{c}#1\\#2\end{array}\right),\left(\begin{array}{c}#3\\#4\end{array}\right)\right]}
\def\Vec#1,#2.{\left(\!\begin{array}{c}#1\\#2\end{array}\!\right)}
\def\vec#1,#2.{{#1\choose{#2}}}
\newcommand{\ket} [1] {\vert #1 \rangle}
\newcommand{\bra} [1] {\langle #1 \vert}

\newcommand{\beq}{\begin{equation}}
\newcommand{\eeq}{\end{equation}}
\newcommand{\beqa}{\begin{eqnarray}}
\newcommand{\eeqa}{\end{eqnarray}}

\begin{document}

\title{Coherent states and the classical-quantum limit considered from the point of view of entanglement.}
\date{}
\author{}
\maketitle
\vglue -1.8truecm
\centerline{Thomas Durt\footnote{Groupe Clart\'e, Institut Fresnel, Domaine Universitaire de Saint-J\'er\^ome,\\ Avenue Escadrille Normandie-Ni\'emen, 13397 \\Marseille Cedex 20, France}Vincent Debierre$^1$ }

\begin{abstract}
Three paradigms commonly used in classical, pre-quantum physics to describe particles (that is: the material point, the test-particle and the diluted particle (droplet model)) can be identified as limit-cases of a quantum regime in which pairs of particles interact without getting entangled with each other.
This entanglement-free regime also provides a simplified model of what is called in the decoherence approach ``islands of classicality'', that is, preferred bases that would be selected through evolution by a Darwinist mechanism that aims at optimising information. We show how, under very general conditions, coherent states are natural candidates for classical pointer states. This occurs essentially because, when a (supposedly bosonic) system coherently exchanges only one quantum at a time with the (supposedly bosonic) environment, coherent states of the system do not get entangled with the environment, due to the bosonic symmetry.
\end{abstract}


\section{Introduction.\label{Intro}}
In order to solve the so-called Quantum Measurement Problem\cite{Measprob}, Zurek and coworkers developped in the framework of the decoherence approach\cite{WikiQDarw,Decoherence1,Decoherence2,ReviewZurek} the idea that, maybe, the classical world emerges from a quantum substrate through an evolutionary process. Zurek proposed an explicit selection process the Environment Induced Selection Rule according to which the islands of stability would correspond to the maximal quantum (Shannon-von Neumann) information\cite{EIN1,EIN2,EIN3}. This approach was baptised ``Quantum Darwinism''\cite{QuantDarw}.

In section~\ref{Factor}, we illustrate this idea in a simple case, the two-body problem in quantum mechanics, and we show that indeed the classical limit of this problem (that corresponds in the framework of the Quantum Darwinist approach to an entanglement free interaction process\cite{Zeit}) is in one to one correspondence with the traditional classical paradigms (point particles, test particle, and diluted particles). Our analysis confirms that, along with Zurek's proposal, the classical islands could be distinguished from the underlying Hilbert space in which they are imbedded according to a principle of maximisation of the relevant information\cite{QD}.

In section~\ref{Coherentstates} we show that under very general conditions (that is, without entering into details at this level, whenever the system coherently exchanges one elementary excitation (``photon'') at a time with the environment), the system's coherent states interact with the environment without getting entangled with it. Consequently, they maximise quantum information and thus play the role of classical pointer states. In particular, this approach makes it possible to estimate in a very simple manner the decoherence rate of ``Schr\"odinger kittens''  similar to those that were prepared by Haroche's team in quantum electrodynamics (QED) cavities in Paris. It also provides a new, highly simplified, manner to derive the Lindblad equation associated to a damping cavity mode.

\section{Classical limit and factorisation-preserving evolution of two interacting quantum particles.\label{Factor}}

\subsection{Quantum Darwinism and Environment Induced selection rule.\label{Darwin}}

Quantum mechanics is astonishingly adequate if we want to describe the material world which we live in. Nevertheless, it is still an open question to know precisely where the border between the microscopic, quantum world and the macroscopic, classical world is situated.

In order to tackle such questions, several conceptual tools have been developed during the three last decades. One of them is the decoherence approach, which provides a general framework aimed at explaining how classicality emerges from a quantum substrate\cite{WikiQDarw,Decoherence1,Decoherence2,ReviewZurek}. As a fine structure in this approach, one distinguishes Quantum Darwinism\cite{QuantDarw}, an approach which implies that during evolution, a selective process isolated in the external, supposedly quantum world, the islands of stability that correspond to the maximal quantum (Shannon-von Neumann) information. This mechanism has been baptised environment induced (EIN) superselection rule\cite{EIN1,EIN2,EIN3}. A key feature of these arguments is the so-called quantum entanglement.

In the following subsections, we introduce the concept of quantum entanglement (section \ref{Entanglement}) and describe results that have relevance in the framework of the
quantum Darwinist approach:
\begin{itemize} 
\item{entanglement is the corollary of interaction\cite{Zeit} (section~\ref{Entint})}
\item{if we apply the criterion of maximal information (section~\ref{Island}) to the simple situation during which two quantum particles interact through a position-dependent potential, in the non-relativistic regime then the classical islands are in one to one correspondence with the three classical paradigms elaborated by physicists before quantum mechanics existed; namely, the droplet or diluted model, the test-particle and the material point approximations\cite{Zeit,QD,Durtlogic} (section~\ref{Decoh}).}
\end{itemize}
       
 \subsection{Entanglement\label{Entanglement}.}
 
The term entanglement was first introduced by Schr\"odinger who described it as the characteristic trait of quantum mechanics, ``the one that enforces its entire departure from classical lines of thought''\cite{Erwin}. Bell's inequalities\cite{EPR,Bell,Bellspeakable} show that when two systems are prepared in an entangled
state, the knowledge of the whole cannot be reduced to the knowledge of the parts, and that to some extent the systems lose their individuality. It is only when systems are not entangled that they behave as separable systems. For instance, it can be shown that whenever two distant systems are in an entangled (pure) state, there exist well-chosen observables such that the associated correlations do not admit a local realist explanation, which is revealed by the violation of well-chosen Bell's inequalities\cite{Gisin,NuovoCimento}. Entanglement thus reintroduces holism and interdependence at a fundamental level and raises the following question: is it legitimate to believe in the Cartesian paradigm (the description of the whole reduces to the description of its parts), when we know that the overwhelming majority of quantum systems are entangled?
   
We shall restrict ourselves in what follows to the simplest case: two systems $A$ and $B$ are prepared in a pure quantum state $\Psi_{AB}$. 
Then the state of the sytem is said to be factorisable at time $t$ whenever the following constraint is satisfied: $\Psi_{AB}\left(t\right)=\psi_{A}\left(t\right)\otimes \psi_{B}\left(t\right)$. Otherwise the system is said to be entangled. When they are in a non-entangled, factorisable, state, subsystems $A$ and $B$ are statistically independent in the sense that the average values of any physical quantity associated to the subsystem $A$ (resp. $B$) is the same that would be obtained if the system $A$ (resp. $B$) was prepared in the state $\psi_{A}\left(t\right)$ (resp. $\psi_{B}\left(t\right)$). Moreover, there are no correlations at all between the subsystems and they can be considered to be independent. 
On the contrary, when subsystems $A$ and $B$ are entangled, the local measurements performed onto them are not statistically independent, that is to say, subsystems $A$ and $B$ exhibit correlations. Those correlations are essentially non-classical in the sense that a classical, separable system would necessarily fulfill statistical constraints called Bell inequalities and that quantum entangled systems violate these constraints\cite{Bell,Gisin,NuovoCimento}.

\subsection{Entanglement and Interaction.\label{Entint}}

Entanglement between $A$ and $B$ is likely to occur whenever they interact\cite{Quote} as is shown by the following property that we reproduce here without proof. The said proof can be found in Ref.~\cite{Zeit}.

Let us consider two interacting quantum systems $A$ and $B$ which obey the Schr\"odinger equation: 
\begin{equation} \label{eq:Schroedinger}
\mathrm{i}\hbar\frac{\partial\Psi_{AB}}{\partial t}=H_{AB}\left(t\right)\Psi_{AB}\left(t\right)\end{equation}
where $H_{AB}\left(t\right)$ is a self-adjoint operator. Then the following property is valid\cite{Zeit}:

{\it A quantum system which intially is in a product state remains in a product state during its evolution (whatever the initial product state is) if and only if the system's Hamiltonian operator can be cast as follows:
\begin{equation} \label{eq:FactHam}
H_{AB}\left(t\right)=H_{A}\left(t\right)\otimes Id._{B}+Id._{A}\otimes H_{B}\left(t\right)\end{equation}
where
$H_{i}$ acts on the $i$th system only while $Id._{j}$ is the identity operator on the $j$th system
($i,j=A,B$).}

In simple words: there is no interaction without entanglement, which establishes that entanglement is very likely to occur in nature. For instance, when we see light coming from a distant star, it is nearly certainly entangled with the atoms that it encountered underway. Entanglement can also be shown to be present in solid structures, {\it e.g.} ferro-magnets and so on, and to play a fundamental role in phase transitions\cite{Osterloh}. 



\subsection{Environment induced superselection rules and classical islands.\label{Island}}

According to the EIN superselection criterion\cite{EIN1,EIN2,EIN3} it can be postulated that, roughly speaking, during evolution, classical islands are preferentially selected thatcorrespond to the minimal increase of Shannon-von Neumann entropy \cite{WikiQDarw,QuantDarw}. These rules correspond to
maximal (Shannon-von Neumann) information. There are various ways to explain the emergence of a classical world that obeys EIN selection rules: 

\begin{itemize}
\item{One can for instance invoke an argument of structural stability: superposition of states belonging to such classical islands would be destroyed very quickly by the interaction with its environment which irretrievably radiates the system's coherence into the environment\cite{ReviewZurek,Omnes}. This process is called ``decoherence'' and is very effective.}
\item Another feature of those islands is their ability to disseminate in the environment copies of their quantum states\cite{zurekmemory}.
\item{It could be on the other side that it is the observer's brain that was led by a slow selection mechanism to ``recognize'' the features of the natural world that contain maximal information; by a Darwinian selection principle the best-informed would be the fittest and consequently would preferentially survive. Extrapolating from Zurek's arguments, one of us (TD) proposed that the classical islands would correspond to the structures that our brain naturally recognises and identifies, and this would explain why the way we think is classical\cite{Zeit,Durtlogic,QD}. We shall come back to this point in the section \ref{discussion}.}

\end{itemize}

\subsection{The decoherence program and the classical limit.\label{Decoh}}

 We applied\cite{Zeit} the quantum Darwinist approach to a very simple situation in which the
system $A$ and its environment $B$ are two distinguishable particles and are described by a (pure) scalar wave function which obeys the non-relativistic Schr\"{o}dinger equation. We also assumed that their interaction potential is an action at a distance that is time-independent and invariant under spatial translations (Coulomb interaction for instance). This is a standard textbook situation which has been studied on countless occasions, for instance in the framework of scattering theory. Systems $A$ and $B$ fulfill thus (in the non-relativistic regime) the following Schr\"{o}dinger equation:
\begin{align} \label{eq:PotSch}
\mathrm{i}\hbar \partial_t\Psi\left(\mathbf{r}_{A},\mathbf{r}_{B},t\right)&=-\frac{\hbar^2}{2}\left(\frac{1}{m_A}\nabla^2_{A}+\frac{1}{m_B}\nabla^2_{B}\right)\Psi\left(\mathbf{r}_{A},\mathbf{r}_{B},t\right)\nonumber\\
&+V_{AB}\left(\mathbf{r}_{A}-\mathbf{r}_{B}\right)\Psi\left(\mathbf{r}_{A},\mathbf{r}_{B},t\right)
\end{align}
where $\nabla^2_{A\left(B\right)}$ is the Laplace operator in the $A\left(B\right)$ coordinates. 
Let us now consider that the system $A$ is the quantum system that interests us, and that the other system is its environment. Actually, the argument is symmetrical, it is indeed clear that this choice is a mere convention.
In order to identify classical islands in this case, we must identify the states that exhibit maximal coherence or maximal Shannon-von Neumann information. We assume here that the full state is pure, which constitutes an oversimplification, because usually interaction with an environment brings decoherence which destroys purity. Nevertheless, as we shall show, one can get interesting insights even in this oversimplified situation.

When a bipartite state $\Psi_{AB}$ is pure, states that obey the EIN selection rule are such that the local or reduced Shannon-von Neumann entropy is minimal (equal to zero) which also means that the reduced states are pure. This only occurs when the full state $\Psi_{AB}$ is a product state, as we explain now. When $\Psi_{AB}$ is pure, it can be written (see section~5.5 of Ref.~\cite{QD}) in the Hilbert-Schmidt (biorthogonal) form: $\ket{\Psi_{AB}}=\sum_{i}\alpha_i\ket{\psi_{A}}^i\ket{\chi_{B}}^i$ where $\left\{\ket{\psi_{A}}^i\right\}$ (resp. $\left\{\ket{\chi_{B}}^i\right\}$) are orthonormal bases of the Hilbert spaces associated to $A$ (resp. $B$). The Shannon-von Neumann entropy is then equal to $S=-\sum_i\left| \alpha_i\right|^2\log \left|\alpha_i\right|^2$, where the Schmidt weights $\left|\alpha_i\right|$ are positive real numbers comprised between 0 and 1 that obey the normalisation constraint $\sum_i\left| \alpha_i\right|^2=1.$ The entropy is strictly positive unless all coefficients are either equal to 0 or equal to 1. By normalisation, only one Schmidt coefficient may be equal to 1 (say the $j^{th}$ one), all others being equal to 0. Then $\ket{\Psi_{AB}}=\ket{\psi_{A}}^j\ket{\chi_{B}}^j$ which is a product state.

In conclusion, two systems in a pure state maximise the Shannon-von Neumann information (minimise the corresponding entropy) whenever their state is factorisable or non-entangled. In other words\cite{EIN3}, {\it ...``Pointer states can be defined as the ones which become
minimally entangled with the environment in the course of the evolution''...}

Then, classical islands correspond to the states that are initially factorisable and remain so throughout their interaction. Actually, in such a case, the particles behave as if they were ``droplets'' (continuous distributions of matter), what corresponds to the so-called 

{\bf Diluted particle regime.}

The regime during which such an entanglement-free evolution occurs can be shown\cite{Zeit} to correspond to what is sometimes called in the literature the mean field or effective field approximation, or the Hartree-Fock approximation\cite{WikiHartreeFock,Gemmer}. In this regime, particles behave as if they were discernable, and constituted of a dilute, continuous medium distributed in space according to the quantum distribution $\left|\psi_{A\left(B\right)}\left(\mathbf{r}_{A\left(B\right)},t\right)\right|^2$. Then, everything happens as if each particle ($A\left(B\right)$) ``felt'' the influence of the other particle as if it were diluted with a probability distribution equal to the quantum value $\left|\psi_{A\left(B\right)}\left(\mathbf{r}_{B\left(A\right)},t\right)\right|^2$. This regime displays similar features to the droplet or diluted particle model that was developed in the pre-quantum period by physicists such as Poincar\'e, Laue, Abraham and others.

{\bf Extreme cases of the diluted particle regime.}

There are two interesting extreme cases of the diluted particle regime: the first one (the test-particle case) corresponds to the case where one of the particles is very heavy relatively to the second particle and well-localised, while the second case (material points) corresponds to a situation where both particles are well localised.

i) {\bf Test-particle regime.}

If the potential only depends on the relative position $\mathbf{r}_{\mathrm{rel}}=\mathbf{r}_{A}-\mathbf{r}_{B}$, if $m_A$ is negligible compared to $m_B$, and providing that the initial state is factorisable and that the $B$ particle is initially at rest and well localised, it can be shown\cite{Zeit} that  the state remains factorisable in time and occupies thus a classical island. This corresponds to what is called the test-particle 
 regime (no feedback of $A$ onto $B$). For instance this is a good approximation of what happens in the hydrogen atom, where the electron is so light that it can be considered to be a test-particle.

ii) {\bf Material points regime.}

Another situation of physical interest is the situation of mutual scattering of two well localised 
wave packets  in the case where we can neglect the quantum extension of
the interacting particles. This will occur when the interaction potential $V_{AB}$ is smooth enough and
particles $A$ and $B$ are described by wave packets with small extension in comparison to the
typical variation length of the potential. It is well known that in this regime, when the de Broglie wavelenghts of the wave packets are small enough, it is consistent to approximate quantum wave mechanics by
its geometrical limit, which is classical mechanics. Indeed, quantum differential cross sections
converge in the limit of small wavelenghts to the corresponding classical cross sections. The Ehrenfest theorem
also predicts that when we can neglect quantum fluctuations, which is the case here, average motions are
nearly classical and provide a good approximation to the behaviour of the full wave-packet in so far we
consider it as a material point. Actually, in this regime, we can replace the interaction potential
by the zeroth order term of its Taylor development around the centers of the wave-packets associated to
particles $A$ and $B$ so that the evolution equation is in good approximation separable into
the coordinates $\mathbf{r}_{A}$ and $\mathbf{r}_{B}$ and we find that\cite{Zeit} if 
\begin{equation*}
\Psi\left(\mathbf{r}_{A}, \mathbf{r}_{B}, t=0\right)=\psi_A\left(\mathbf{r}_{A}, t=0\right)\psi_B\left(\mathbf{r}_{B}, t=0\right)
\end{equation*}
then
\begin{equation*}
\Psi\left(\mathbf{r}_{A}, \mathbf{r}_{B}, t\right)\approx \psi_A\left(\mathbf{r}_{A}, t\right)\psi_B\left(\mathbf{r}_{B}, t\right).
\end{equation*}

\section{Coherent states reconsidered in the Quantum Darwinist approach\label{Coherentstates}.}

Let us assume that the system and the environment are made of bosons. Moreover, we shall assume that there exists an orthonormal basis on which we can expand the states of the system (resp. environment) that is labelled by the number $n^S$ (resp. $n^E$) of ``elementary excitations'' of the modes of the bosonic field that characterises the system (resp. environment). Such elementary excitations are for instance so-called photons in the case where we consider the state of a certain mode of the electro-magnetic field inside a QED cavity, like in experiments carried out by Haroche and his team, but they could be phonons, excitons and so on in other situations. Our results are also valid in first quantization, in the case of massive particles as we show elsewhere\cite{rapportdestage}. A pure state $\ket{\Psi_S}$ ($\ket{\Psi_E}$) of the system (environment) can always be expressed as a coherent superposition of energy states (Fock states in the case of light):
\begin{equation} \label{eq:FockSE}
\ket{\Psi_S}=\sum^{+\infty}_{n=0}\psi^S_n\ket{n_S}\,\,\,\,\,\left(\ket{\Psi_E}=\sum^{+\infty}_{m=0}\psi^E_m\ket{m_E}\right).
\end{equation}
Our last assumption is that the system coherently exchanges elementary excitations with its environment, that is, if at time $t=0$,
\begin{equation*}
\ket{\Psi_{S-E}\left(t=0\right)}=\ket{1_S}\otimes\ket{0_E}
\end{equation*}
then at time $t$,
\begin{equation}
\ket{\Psi_{S-E}\left(t\right)}=\alpha\left(t\right)\ket{1_S}\otimes\ket{0_E}+\beta\left(t\right)\ket{0_S}\otimes\ket{1_E}\label{ansatz}
\end{equation}
where $\alpha\left(t\right)$ and $\beta\left(t\right)$ are properly normalised complex amplitudes (that is, they verify $\left|\alpha\left(t\right)\right|^2+\left|\beta\left(t\right)\right|^2=1$) that we assume to know ({\it e.g.} through a Wigner-Weisskopf approach to the problem). In principle the exact time-dependence of $\alpha$ and $\beta$ can be derived once we know the interaction Hamiltonian $H_{SE}$ between the system ($S$) and its environment ($E$). We shall give two explicit examples in the following.

It is worth noting that without losing generality, an energy state is superposition of different (sometimes more than one) modes. In cavity QED experiments, the system (light inside the cavity) is characterized by a single mode, but the environment is associated to an infinity of modes.

Since we consider undistinguishable particles (here, bosons), we can infer how a state which initially consists of $n$ elementary excitations of the system and zero excitation of the environment (that is, at $t=0$, the environment is prepared in its vacuum state) will evolve. We get
\begin{align} \label{eq:Evolution}
\ket{\Psi_{S-E}\left(t=0\right)}&=\ket{n_S}\otimes\ket{0_E}\nonumber\\
\rightarrow\ket{\Psi_{S-E}\left(t\right)}&=\sum^{n}_{m=0}\sqrt{\frac{n!}{m!\left(n-m\right)!}}\alpha^m\left(t\right)\beta^{n-m}\left(t\right)\ket{m_S}\otimes\ket{\left(n-m\right)_E}
\end{align}
where the square root comes from symmetrisation over all Fock states having $m$ particles in the system and $n-m$ in the environment. In particular, if at time $t=0$ the system is prepared in a coherent state and the environment is in the vacuum state, then the evolution yields coherent states for the environment:
\begin{align} \label{eq:Evocoh}
\ket{\Psi_{S-E}}\left(t=0\right)&=\mathrm{e}^{-\frac{\left|\lambda\right|^2}{2}}\sum^{+\infty}_{n=0}\frac{\lambda^n}{\sqrt{n!}}\ket{n_S}\otimes\ket{0_E}\nonumber\\
\rightarrow\ket{\Psi_{S-E}}\left(t\right)&=\mathrm{e}^{-\frac{\left|\lambda\right|^2}{2}\left|\alpha\left(t\right)\right|^2}\sum^{+\infty}_{n=0}\sum^{n}_{m=0}\frac{\left(\lambda\alpha\left(t\right)\right)^m}{\sqrt{m!}}\ket{m_S}\otimes\mathrm{e}^{-\frac{\left|\lambda\right|}{2}\left|\beta\left(t\right)\right|^2}\frac{\left(\lambda\beta\left(t\right)\right)^{n-m}}{\sqrt{\left(n-m\right)!}}\ket{\left(n-m\right)_E}
\end{align}
where we used the identity $\left|\alpha\left(t\right)\right|^2+\left|\beta\left(t\right)\right|^2=1$. 

 The state at time $t$ can be rewritten as a product of coherent states:
 
 \begin{align}&\ket{{\bf \Psi}_{S-E}( t=0)}=\nonumber\\&e^{-{|\lambda|^2\over 2}|\alpha(t)|^2}\Sigma^{\infty}_{n-m=0}{(\lambda\alpha(t))^{(n-m)}\over \sqrt{(n-m)!}}\ket{(n-m)_{S}}\otimes e^{-{|\lambda|^2\over 2}|\beta(t)|^2}\Sigma^{\infty}_{m=0}{(\lambda\beta(t))^m\over \sqrt{m!}}\ket{m_{E}}\label{coherent}\end{align}
  
  This establishes that coherent states of the system, in this regime, interact with the environment without getting entangled with it. They can thus be considered as ``classical pointers'' according to the criterion for classicality derived by Zurek in the framework of the quantum Darwinist approach (see sections \ref{Island} and \ref{Decoh}).
  
  This is a well-known result but the originality of our approach is to explain it in terms of the bosonic symmetry of the system and of the environment and of the ansatz (\ref{ansatz}). This ansatz, in turn, expresses the unitarity of the coupling to the environment.

\subsection{Schr\"odinger kittens.\label{Meowth}}

In particular, if at time $t=0$ the state of the system is a coherent and symmetrical superposition of coherent states of opposite parity (what is called Schr\"odinger kittens in the literature), while the environment is prepared in the vacuum state:
\begin{equation} \label{eq:KittenBirth}
\ket{\Psi_{S-E}\left(t=0\right)}=\frac{\mathrm{e}^{-\frac{\left|\lambda\right|^2}{2}}}{\sqrt{2\left(1+\mathrm{e}^{-2\left|\lambda\right|^2}\right)}}\left(\sum^{+\infty}_{n=0}\frac{\lambda^n}{ \sqrt{n!}}\ket{n_S}+\sum^{+\infty}_{n=0}\frac{\left(-\lambda\right)^n}{ \sqrt{n!}}\ket{n_S}\right)\otimes\ket{0_E}
 \end{equation}
then at time $t$ the state of the full system will be given by
\begin{align} \label{eq:KittenGrowth}
\ket{\Psi_{S-E}\left(t\right)}&=\frac{\mathrm{e}^{-\frac{\left|\lambda\right|^2}{2}}}{\sqrt{2\left(1+\mathrm{e}^{-2\left|\lambda\right|^2}\right)}}\left[\sum^{+\infty}_{n=0}\sum^{n}_{m=0}\frac{\left(\lambda\alpha\left(t\right)\right)^m}{\sqrt{m!}}\ket{m_S}\otimes\frac{\left(\lambda\beta\left(t\right)\right)^{n-m}}{ \sqrt{\left(n-m\right)!}}\ket{\left(n-m\right)_E}\right.\nonumber\\
&\left.+\sum^{+\infty}_{n=0}\sum^{n}_{m=0}\frac{\left(-\lambda\alpha\left(t\right)\right)^m}{\sqrt{m!}}\ket{m_S}\otimes\frac{\left(-\lambda\beta\left(t\right)\right)^{n-m}}{ \sqrt{\left(n-m\right)!}}\ket{\left(n-m\right)_E}\right].
 \end{align}
One way to quantify the decoherence of the system is to estimate the interference between the two ``kittens'' (that is between the two coherent states of opposite parity).  Therefore we must firstly estimate the reduced kitten state
\begin{equation*}
\rho_S\left(t\right)=\mathrm{Tr}_E\ket{\Psi_{S-E}\left(t\right)}\bra{\Psi_{S-E}\left(t\right)}.
\end{equation*}
This reduced state, at time $t$, obeys
\begin{subequations}
\begin{align}
\rho_S\left(t\right)&=\frac{\mathrm{e}^{-\left|\lambda\alpha\left(t\right)\right|^2}}{2\left(1+\mathrm{e}^{-2\left|\lambda\right|^2}\right))}\sum^{+\infty}_{n=0}\frac{\left(\lambda\alpha\left(t\right)\right)^n}{ \sqrt{n!}}\ket{n_S}\sum^{+\infty}_{k=0}\frac{\left(\lambda\alpha\left(t\right)\right)^k}{ \sqrt{k!}}\bra{k_S}\label{un}\\
&+\frac{\mathrm{e}^{-\left|\lambda\alpha\left(t\right)\right|^2}}{2\left(1+\mathrm{e}^{-2\left|\lambda\right|^2}\right)}\sum^{+\infty}_{n=0}\frac{\left(-\lambda\alpha\left(t\right)\right)^n}{\sqrt{n!}}\ket{n_S}\sum^{+\infty}_{k=0}\frac{\left(-\lambda\alpha\left(t\right)\right)^k}{\sqrt{k!}}\bra{k_S} \label{deux}\\
&+\mathrm{e}^{-2\left|\lambda\beta\left(t\right)\right|^2}\frac{\mathrm{e}^{-\left|\lambda\alpha\left(t\right)\right|^2}}{2\left(1+\mathrm{e}^{-2\left|\lambda\right|^2}\right)}\sum^{+\infty}_{n=0}\frac{\left(\lambda\alpha\left(t\right)\right)^n}{ \sqrt{n!}}\ket{n_S}\sum^{+\infty}_{k=0}\frac{\left(-\lambda\alpha\left(t\right)\right)^k}{ \sqrt{k!}}\bra{k_S}\label{trois}\\
&+\mathrm{e}^{-2\left|\lambda\beta\left(t\right)\right|^2}\frac{\mathrm{e}^{-\left|\lambda\alpha\left(t\right)\right|^2}}{2\left(1+\mathrm{e}^{-2\left|\lambda\right|^2}\right)}\sum^{+\infty}_{n=0}\frac{\left(-\lambda\alpha\left(t\right)\right)^n}{\sqrt{n!}}\ket{n_S}\sum^{+\infty}_{k=0}\frac{\left(\lambda\alpha\left(t\right)\right)^k}{\sqrt{k!}}\bra{k_S}\label{quatre}
\end{align}
\end{subequations}
 
\subsection{Decoherence between the kittens.\label{KitKat}} 
We shall assume that $\left|\lambda\right|^2$ is larger than, say, 5, sothat the two kittens are in a good approximation orthogonal ($\mathrm{e}^{-2\left|\lambda\right|^2}\leq \mathrm{e}^{-50}$).

The first two contributions (\ref{un},\ref{deux}) correspond to an incoherent superposition of the two kitten states, while the two last contributions (\ref{trois},\ref{quatre}) are interference terms.  As we see here, the interference is damped by a factor $\mathrm{e}^{-2\left|\lambda\beta\right|}$, which indicates that the coherence time decreases exponentially with $\left|\lambda\right|$. 

In principle, it is sufficient that one elementary excitation of the system at time 0 gets transfered to the environment in order that one can distinguish between the two kittens. For instance a macroscopic superposition of a same excited atom over two distant locations of space can be broken in principle once this atom emits only one photon. This explains intuitively why the coherence time decreases exponentially with $\left|\lambda\right|$ which is the average number of elementary excitations of the system at time 0. 
 
 Actually, it is more general to require that decoherence occurs when the the two kittens states are in good approximation equal to the Schmidt basis of the system, for a Schmidt number equal to two. Then the state of the system can be written as follows:

 \begin{equation} 
\ket{\Psi_{S-E}\left(t\right)}\approx {1\over\sqrt{2}}\left[\sum^{+\infty}_{l=0}\frac{\left(\lambda\alpha\left(t\right)\right)^l}{\sqrt{l!}}\ket{l_S}\otimes \ket{+_E}\nonumber\\
+ \sum^{+\infty}_{l=0}\frac{\left(-\lambda\alpha\left(t\right)\right)^l}{\sqrt{l!}}\ket{l_S}\otimes\ket{-_E} \right].
 \end{equation}

where the environment states  $\ket{+_E}$ and $\ket{-_E}$ are orthogonal in good approximation. They play then the role of record states, which means that the interaction has disseminated the information relative to the pointer states into the environment. In our case this occurs when $\left|\lambda\beta\right|^2$ gets larger than, say, 5. In other words, we need the exchange of a few photons to make sure that the environment accurately records the kitten states. One photon does not suffice, due to the overlap of the ``kitten'' coherent states which is an intrinsic feature of our model.

Then, by performing a measurement of a non-degenerate observable diagonal in the $\ket{+_E}$ and $\ket{-_E}$ basis, it is possible in principle to distinguish between them, and, making use of the correlations between the system and its environment (which are prepared in a maximally entangled Bell state in this case), one could infer in principle to which kitten state the system belongs.
 
Due to the no-signaling theorem, the reduced state of the system is the same regardless of whether we perform this hypothetical measurement, which explains ultimately why coherence is destroyed between the (system) kitten states. 
 
In the framework of the decoherence approach this is a common way to tackle the measurement problem.\cite{ReviewZurek} Of course, it can be argued that, although this approach wonderfully explains the decrease of the interferences exhibited by the system after tracing out the environment, it cannot explain why a particular kitten state is observed whenever we observe the system\cite{decoSchlos} (we shall come back to this point in the section \ref{discussion}).
 
Two interesting extreme cases appear, in the case that (i) the interaction between the system and its environment is fully coherent (this would correspond to the so-called quantum eraser, when the system's coherence is recovered after a while), and (ii) the interaction between the system is dissipative but markovian (that is, the survival probability of an elementary excitation inside the cavity decreases exponentially with time, as is the case in QED cavity experiments).
 
 \subsection{Fully coherent interaction between the system and its environment.\label{Regimecoh}}
 
The fully coherent regime is reached for instance in the simplest case, when two coupled oscillators (modes) interact through the interaction Hamiltonian
\begin{equation*}
H_{ES}=\mathrm{i}\lambda\hbar\left(a_E^{\dagger}\,a_S-a_E\,a_S^{\dagger}\right)
\end{equation*}
(written in function of the creation and annihilation phonon operators of the oscillators $E$ and $S$).

The full Hamiltonian of the system constituted by $E$ and $S$ is then\cite{Durtlogic}
\begin{equation} \label{eq:Hamilton}
H=H_E+H_S+H_{ES}=\hbar\left[\omega_Ea_S^{\dagger}\,a_S+\omega_Sa_E^{\dagger}\,a_E+\mathrm{i}\hbar\kappa\left(a_E^{\dagger}\,a_S-a_E\,a_S^{\dagger}\right)\right]
\end{equation}
where $\omega_{E\left(S\right)}$ are the oscillator frequencies and $\kappa$ is the coupling constant between these oscillators. Energy conservation imposes that $\omega_S=\omega_E=\omega$.

It is easy to check in this case that  the states $\ket{1_S}\otimes\ket{0_E}\pm i \ket{0_S}\otimes\ket{1_E}$  are eigenstates of the Hamiltonian for the energies $\omega\pm \kappa$.

Then, the state of the full system precesses between the states $\ket{1_S}\otimes\ket{0_E}$ and $\ket{0_S}\otimes\ket{1_E}$ so that $\alpha\left(t\right)$$=\mathrm{e}^{i\omega t}\cos\left(\kappa t\right)$ and $\beta\left(t\right)=\mathrm{e}^{i\omega t}\sin\left(\kappa t\right)$. Due to the process of coherent regeneration of the initial state after a period ${2\pi\over \kappa}$, the decoherence induced by the environment gets periodically erased, and restored a  quarter of period later. After a half-period, the states of the environment and of the system are swapped.

\subsection{Dissipative and memory-free (Markovian) interaction between the system and its environment.\label{Regimedis}}

\subsubsection{A Wigner-Weisskopf approach.\label{WW}}

The dissipative regime is reached for instance when the environment is coupled to an infinity of oscillators that constitute the environment (supposedly prepared initially at temperature 0). Most commonly, the distribution of oscillators is discrete\cite{bookharoche} but a coupling to a continuum of oscillators leads to the same results\cite{rapportdestage}. In this case, coherence gets irreversibly dissipated in the environment, and $\left|\alpha\left(t\right)\right|^2=\mathrm{e}^{-\Gamma t}$, where $\Gamma$ is the loss-rate of the system (for instance the dissipation rate of a particular mode of the electro-magnetic field inside a QED cavity). 

For small times, $\left|\beta\left(t\right)\right|^2$ increases linearly with time (which corresponds to the Fermi golden rule, valid in the Wigner-Weisskopf approach which typically describes this type of behaviour), and we see that the coherence decreases like $\mathrm{e}^{-\Gamma \left|\alpha\left(t=0\right)\right|^2 t}$ so that the coherence time is inversely proportional to $\Gamma \left|\alpha\left(t=0\right)\right|^2$, which corroborates our discussion of section \ref{KitKat}.

 \subsubsection{Derivation of the Linblad master equation \it{\`a la} Monte-Carlo.\label{Lindblad2}}

A convenient way to derive the Lindblad master equation is to introduce the so-called quantum Monte-Carlo trajectories that are such that, after averaging over many of them we recover the exact, full, density matrix of the reduced state of the system. In the present case, where the situation is very simple, it is not too difficult to guess the right Monte-Carlo process. Indeed, decoherence, dissipation and so on are characterized by a unique factor, $\Gamma$, the loss-rate of the system. Let us assume that at time $t$ the system is a $n$ photon Fock state (more generally an energy state that consists of $n$ elementary excitations). During the interval $[t,t+dt]$, it is thus likely that one elementary excitation of the system will get dissipated in the environment, with probability $\Gamma n dt$, in which case the state at time $t$ ought to be replaced by the properly normalised $(n-1)$ photon Fock state ${a\Psi_S(t)\over \sqrt n}$ at time $t+dt$. Otherwise (and this happens with probability $1-\Gamma dt$), the state of the system at time $t+dt$ ought to be equal to $\sqrt{1-\Gamma n dt }(\Psi_S(t)+{H\over i\hbar}dt\Psi_S(t)).$

In average, $\rho(t)$ evolves thus during the time interval $[t,t+dt]$ to $\rho(t+dt)=\rho(t)+{\Gamma n a\rho(t) a^\dagger\over n}+dt[{H\over i\hbar},\rho(t)]-{\Gamma n dt\over 2}\rho(t)$, that is,

\begin{equation}{d\rho(t)\over dt}=[{H\over i\hbar},\rho(t)] +{\Gamma\over 2}( 2 a\rho(t) a^\dagger-a^\dagger a
 \rho(t)-\rho(t)a^\dagger a).\label{master}\end{equation}
 
 \subsubsection{A simple derivation of the Linblad master equation.\label{Lindblad}}

 This is not the unique way to derive the master equation that describes the relaxation of a cavity field in an environment at zero temperature, but it is certainly one of the simplest methods. Here, we shall show that in our approach the derivation of the master equation is even simplified.
 The basic ingredient of our derivation is to note that each damped coherent state of the form 
\begin{equation}\ket{\Psi_{S}}\left(t\right)=\mathrm{e}^{-\frac{\left|\lambda\right|^2}{2}\left|\alpha\left(t\right)\right|^2}\sum^{+\infty}_{m=0}\frac{\left(\lambda\alpha\left(t\right)\right)^m}{\sqrt{m!}}\ket{m_S},\end{equation}with $\alpha(t)=\mathrm{e}^{i\omega t-{\Gamma\over 2}t}$, and $H=\hbar \omega a^\dagger a$, obeys the Lindblad equation (\ref{master}). Now this is the reduced state of the state that satisfies equation (\ref{coherent}), and this property is true for any complex value of $\lambda$. As coherent states constitute a basis of the Hilbert space, any state (pure or mixed) must obey the master equation (\ref{master}), by linearity of the master equation. This ends our derivation.
\subsection{Classical limit.}In many standard text-books, the discussion of the classical limit reduces to the presentation of Ehrenfest's theorem. Actually, the problem is more subtle than it could appear at first sight. One of the reasons therefore is that Ehrenfest's theorem goes about unitary, {\it \`a la} Schr\"odinger evolution.

In no-collapse theories, it makes sense to consider observables without observers, and to assume that the evolution of the ``wave function of the universe'' (or, more modestly of the system of interest plus its direct environment) is unitary but if we consider subsystems unitarity is no longer guaranteed. In collapse theories it is difficult {\it per se} to avoid departures from unitarity, because the collapse process is not unitary. 

Another problem is that Ehrenfest's theorem also goes about average values, which presupposes that probabilities are assigned to the physical quantities that characterize the system, but it is not an easy task to infer Born's rule from no collapse theories (unless it is assumed to hold from the beginning like in the Bohmian interpretation). 

Besides, it is not clear how a classical picture could emerge from a quantum substrate, where entanglement is the rule as we discussed in section \ref{Entint}. Regarding this problem, it is worth noting that an interesting property of pointer states is that they do not get entangled with the environment. As we emphasised in section \ref{Decoh}, entanglement being non-classical in essence,  due to non-locality which has no classical counterpart, it is thus a relevant criterion of classicality to impose that a classical system disentangles from the environment.

This justifies why, leaving aside these problems of interpretation, we shall from now on formulate the problem of the classical limit in the following restricted sense: 

``do classical pointer states behave like classical systems?''

 This question deserves to be studied by itself. As we know that, by definition, coherent superpositions of classical pointer states  will vanish very quickly, we can infer that, in the case that classical pointer states behave like classical systems, the system will behave in average like a mixture of classical states. In particular, classical ``pure'' states will correspond to pure pointer states. 

In the case under study in this section, classical pointer states are coherent states and we must answer to the question:

``do coherent states behave like classical systems?''

The answer is yes, for several reasons:

-As was noted by Schr\"odinger, coherent states behave like classical systems, because their quantum fluctuations are minimal in the sense that they saturate Heisenberg's uncertainties. Moreover these fluctuations remain the same when the average number of elementary excitations increases. In the high-energy limit they can thus consistently be neglected. 

-Another interesting feature of coherent states is that their physics (in absence of dissipation) is the same as for an oscillator, that is, they undergo (when expressed in convenient units and parameters) a linear force field. In this case, the average value of the force is the force estimated in the average value of the position, so that Ehrenfest's theorem predicts that the average position exactly obeys Hamilton's equation. For situations where the force does not linearily depend on the position, quantum fluctuations will always induce a departure from Hamilton's equations.

-In the limit of high energy coherent states, it is thus likely that in good approximation Maxwell's equations will be valid, because they are known to correspond to an infinity of classical decoupled oscillators.

-In the case of a QED cavity, discussed in section \ref{Regimedis}, the system behaves as an oscillator coupled to its environment and it obeys a non-unitary master equation. Nevertheless, one can check that its average values exactly obey the same equation of motion as a damped classical oscillator. Intuitively, the present discussion justifies why it makes sense to consider that the dissipation factor of a QED cavity, $\Gamma$, can be taken to be equal to the classical dissipation factor. This is so because the correspondence principle is fully justified in the case of coherent states, even in the case of damped coherent states.

\section{Interpretation of the EIN selection rule.\label{discussion}}
Retrospectively, it is tempting to reconsider the first motivation of the decoherence approach and of the quantum Darwinist approach in particular which was to solve the measurement problem\cite{Measprob}.
Decoherence (see Refs.\cite{decoSchlos} and \cite{revZurek2} for a review) is an essential ingredient of all the so-called no collapse interpretations (many worlds, Bohmian interpretation and so on). It is also a mechanism that allows us to understand why quantum interferences are so rare and difficult to observe at the macroscopic scale. It is thus a useful and powerful tool, even in interpretations where a quantum-classical duality is supposed to be present from the beginning (this concerns spontaneous localisation models but also, to some extent, the Copenhagen interpretation, and approaches {\it \`a la} Leggett where a border line supposedly exists that separates the quantum and classical domains). In our view, the EIN selection rule has the merit to provide a very convenient strategy for tackling the question of the classical limit, as hopefully was illustrated by the results of the previous sections.

We doubt that the EIN selection {\it per se} suffices to the resolution of the measurement problem, because in the mere definition of ``information'', the notion of probability plays a crucial role, and probability is at the heart of the quantum measurement problem. In other words, information is defined in terms of statistical distributions of what John Bell called ``beables''\cite{Bellspeakable}, that is, classical, realistic quantitities. Dissimulating these beables under a mask constituted by information-theoretic concepts does not really solve the problem.

Now, leaving aside the measurement problem, it is possible to interpret differently the status of the EIN as we discuss now. One of us (TD) proposed that these classical islands would correspond to the structures that the observer naturally recognises and identifies, and this would explain why the way we think is classical\cite{Zeit,Durtlogic,QD}.  It is implicit in our approach that, if a quantum Darwinist selection principle of this type sculpted the way living creatures see the outside world, this could only occur through a repeated series of measurements through which these creatures interacted with their environment. During such measurements, potentialities get actualised, which is the main issue lying at the heart of the measurement problem, as is illustrated by Omn\`{e}s's zeroth principle\cite{Omnes}: ``reality exists''. Hence there remains, in our view, an intrinsic loophole in the quantum Darwinist approach (in the case that no extra-argument such as the many world interpretation is enforced into the discussion), a loophole that we did not intend to solve here.

Note that many explanations of the world in terms of evolutionary processes suffer from similar loopholes. For instance, an ``explanation'' of the odd dimensionality of space based on the Huygens principle (sketeched in appendix \ref{huygens}) that we outlined in Ref.\cite{Durtlogic} also suffers from some circularity,  what is illustrated by the paradoxical sentence {\it ``The world has three dimensions because we listen to music''}. 

Despite of this intrinsic limitation, we consider that it can be enlightening and useful to ask whether nature obeys a principle of optimality. Such principles played an important role in modern physics, and were at the source of fruitful discussions and research, as illustrated by Maupertuis's least action principle...

In appendix \ref{Batman}, we give an example of an evolutionary process based on information. Bat signals have been shown to optimise the information gain. They possess a non-trivial Fourier signature, which confirms that in Nature Darwinist selection processes can, in certain circumstances, privilege strategies aimed at optimising information.

\section{Conclusions.\label{Ccl}}

In this paper, we collected a series of results that appeared in several publications\cite{Zeit,QD,Durtlogic}, and discussed the evolution of coherent states. The key idea underlying our analysis is that, in accordance with the quantum Darwinist approach, a principle of maximal information could explain how the classical world emerges from a quantum substrate. 

In the case of two interacting particles, in a non-relativistic, first quantised treatment, {\it \` a la} Schr\"odinger, the EIN principle allows us to identify three classical paradigms: the diluted particle or droplet model, and its two extreme limit cases, the test particle regime and the material points regime. 

In particular, the idea to associate classical pointer states to the states that maximise Shannon-von Neumann entropy sheds a new light on the concept of coherent states. As we have shown in the section \ref{Coherentstates}, under very general conditions, coherent states are classical pointer states. Moreover, it is sufficient to know the rate of coherent transfer of ``photons'' (here seen as elementary excitations of the modes of the field that we are dealing with) between the system and the environment in order to be able to estimate the decoherence of the system that results from its interaction with the environment. In particular, in the case of ``Schr\"odinger kittens'' which has been studied by Haroche's team, we are able to estimate precisely the rate of coherence loss induced by the environment on the basis of a minimal set of assumptions about the interaction between the system and its environment, by direct computation and without resorting to more sophisticated techniques (like the Lindblad equation, the quantum Monte-Carlo approach and so on).

The concept of classical pointer state can thus be seen to provide a useful tool for tackling the problem of decoherence, and also the classical limit.

\section{Appendices}
\subsection{How the Huygens principle selects spaces of odd dimensionality.\label{huygens}}
Interestingly, the study of wave propagation makes it possible to connect the dimensionality of space-time and the Huygens principle, on the basis of classical information concepts.\cite{Huygens,Ehrenfest1,Ehrenfest2} This is not so much amazing in a sense, because wave propagation is intimately related to propagation of information at the classical level, in other words of communication, another aspect of information\footnote{ Claude Shannon wrote hereabout {\it ''The fundamental problem of communication is that of reproducing at one
point either exactly or approximately a message selected at another point''} in his famous paper ``A mathematical theory of communication''\cite{shannon}.}. The connection between Huygens principle and the dimension of space relies on a study due to Hadamard at the beginning of the 20th century concerning the concept of dimension (see Ref.\cite{Huygens} and references therein): in a space-time of d+1 dimensions, wave propagation ruled by d'Alembert's equation obeys Huygens principle whenever $d$ is an odd and positive integer. This claim can be translated into informational terms\cite{Ehrenfest1,Ehrenfest2}: Huygens principle is satisfied when the state of the image reproduces accurately the state of the source (after a time delay proportional to the distance between image and source). It is this property that allows us to obtain a fidel representation of the (3+1) external world by using sensitive receptors such as our ears and our eyes. In dimension 2+1 for instance Huygens principle is not satisfied. Adopting an information theoretic approach, one could invert the reasoning and assume that maybe the conventional 3+1 representation of space-time was privileged because it is informationally advantageous to do so. It could occur indeed that other physical phenomena that are characterised by other dimensions coexist but that we are blind to them simply because they are informationally deprived of interest and of sense. Considered so, the dimension of space would no longer be absolute but it would rather be a consequence of an evolutionary process that induced us to select a representation of the world that is advantageous and optimal from the point of view of information.

The next section provides an explicit example showing that such processes are not pure fantasy. 

\subsection{Example of an evolutionary process in which the best-informed survives: about bats, chirps and maximising classical information.\label{Batman}} 

The sonars of bats are sophisticated engines, and, rather surprisingly, the ultrasonic waves that they send in order to explore their environment are highly non-trivial\cite{Max}. 
Actually, these signals belong to the class of so-called ``chirp'' impulsions, which have been shown, in the framework of radar technology, to optimise the information gain. At the beginning of the development of human made radars, the signals sent by the radars possessed a ``trivial'' ``ping-pong'' like signal, that is, it consisted of impulsions peaked in time, similar to Dirac deltas. Chirp signals possess a non-trivial Fourier signature that definitively differentiates them from ``ping-pong'' like signals. They were found to provide an optimal solution to the radar detection several years after the development of the first radars. We do not want to comment this result in detail but our main remark is the following: the recognition of the advantages of chirp impulsions at the level of human made radar technology was  only made possible thanks to the development of modern information theory. It is remarkable that, through an evolutionary process, bats were also led to adopt the most efficient, and highly non-trivial strategy that consists of sending chirp signals instead of peaked impulsions. More, they adopted this strategy millions of years before man did. Considered so, bats provide a living confirmation that in Nature Darwinist selection processes can, in certain circumstances, privilege strategies aimed at optimising information.

\end{document}